\documentclass[twocolumn]{emulateapj}

\newcommand{\adhoc}{{\tt ADHOCw}}
\newcommand{\FP}{Fabry-Perot}
\newcommand{\fantomm}{{\tt FaNTOmM}}

\newcommand{\Ha}{H$\alpha$}
\newcommand{\Vlos}{$V_\mathrm{los}$}
\newcommand{\Vsys}{$V_\mathrm{sys}$}

\newcommand{\Op}{$\Omega_\mathrm{p}$}
\newcommand{\Opp}{$\Omega_\mathrm{p}^\mathrm{P}$}
\newcommand{\Ops}{$\Omega_\mathrm{p}^\mathrm{S}$}
\def\kms{$\mbox{km s}^{-1}$}
\def\kmskpc{$\mbox{km s}^{-1}\mbox{ kpc}^{-1}$}

\def\deg{^\circ}

\def\farcs{\hbox{$.\!\!^{\prime\prime}$}}

\shorttitle{Quantifying Resonant Structure in NGC~6946}
\shortauthors{Fathi et al.}

\slugcomment{Accepted for publication in ApJ Letters (Aug 1st 2007)}
\begin{document}

\title{Quantifying Resonant Structure in NGC~6946 from Two-dimensional Kinematics}
\author{Kambiz Fathi\altaffilmark{1,2}, Silvia Toonen\altaffilmark{1,3}, Jes\'us Falc\'on-Barroso\altaffilmark{4}, John E. Beckman\altaffilmark{1,5}, Olivier Hernandez\altaffilmark{6}, Olivier Daigle\altaffilmark{6}, Claude Carignan\altaffilmark{6}, Tim de Zeeuw\altaffilmark{3}}

\altaffiltext{1}{Instituto de Astrof\'\i sica de Canarias, C/ V\'\i a L\'actea s/n, 38200 La Laguna, Tenerife, Spain. Email: fathi@iac.es; jeb@iac.es}
\altaffiltext{2}{Stockholm Observatory, AlbaNova University Center, 106 91 Stockholm, Sweden}
\altaffiltext{3}{Sterrewacht Leiden, University of Leiden, Niels Bohrweg 2, 2333 CA Leiden, The Netherlands. Email: silvia.toonen@gmail.com; tim@strw.leidenuniv.nl}
\altaffiltext{4}{European Space Agency / ESTEC, Keplerlaan 1, 2200 AG Noordwijk, The Netherlands. Email: jfalcon@rssd.esa.int}
\altaffiltext{5}{Consejo Superior de Investigaciones Cient\'\i ficas, Spain}
\altaffiltext{6}{Universit\'e de Montr\'eal, C.P. 6128 succ. centre ville, Montr\'eal, QC, Canada H3C 3J7. Email: olivier@astro.umontreal.ca; odaigle@astro.umontreal.ca; carignan@astro.umontreal.ca}

\begin{abstract}
We study the two-dimensional kinematics of the \Ha-emitting gas in the nearby barred Scd galaxy, NGC~6946, in order to determine the pattern speed of the primary $m=2$ perturbation mode. The pattern speed is a crucial parameter for constraining the internal dynamics, estimating the impact velocities of the gravitational perturbation at the resonance radii, and to set up an evolutionary scenario for NGC~6946. Our data allows us to derive the best fitting kinematic position angle and the geometry of the underlying gaseous disk, which we use to derive the pattern speed using the Tremaine-Weinberg method. We find a main pattern speed \Opp=$22^{+4}_{-1}$ \kmskpc, but our data clearly reveal the presence of an additional pattern speed \Ops=$47^{+3}_{-2}$ \kmskpc\ in a zone within 1.25 kpc of the nucleus. Using the epicyclic approximation, we deduce the location of the resonance radii and confirm that inside the outer Inner Lindblad Resonance radius of the main oval, a primary bar has formed rotating at more than twice the outer pattern speed. We further confirm that a nuclear bar has formed inside the Inner Lindblad Resonance radius of the primary bar, coinciding with the inner Inner Lindblad Resonance radius of the large-scale $m=2$ mode oval.\looseness-2 
\end{abstract}


\section{Introduction}
Secular evolution of structure in disk galaxies is largely governed by internal gravitational processes. In this context, density waves play an important role as a driver for the internal evolution. Self-gravitating bars contain a significant fraction of the disk material and thus have a prominent role in the evolution of their host disks. Bars are efficient drivers for redistribution of angular momentum within their host galaxies, leading to mass transfer from the outer parts towards the circumnuclear regions (e.g., Shlosman, Frank, \& Begelman 1989), ultimately via structures such as  nuclear bars or spirals, continuing to the vicinity of their host galaxy's nuclear supermassive black hole (e.g., Fathi et al. 2006). An increasing number of observational as well as theoretical results suggest that secular evolution of disks due to density waves does occur, but the details on how the structures and the internal dynamics are affected require more inputs, in particular to constrain the mass models in the inner parts of galaxies where the rotation curve is still rising. \looseness-2 

A key parameter in self-consistent models of barred galaxies is the angular rate at which the bar pattern rotates, i.e., the pattern speed of the bar, denoted by \Op. This parameter is needed to determine whether a bar is spontaneous or tidal, to derive the velocity fields as well as mass transfer rates, and hence the densities, in perturbed disks, or to establish whether spiral arms and any secondary bars are driven by resonance interactions with primary bars (e.g., Shlosman \& Heller 2002; Block et al. 2004). However it has proved complex to determine this parameter observationally. \looseness-2 

The most common technique is based on assumed knowledge of the locations of the resonances (Buta \& Combes 1996) which in turn involves understanding the behavior of stars and gas at these resonances. Thus, purely observationally, \Op\ can be estimated by the locations of the resonances using surface brightness profiles of galaxies (e.g., Puerari \& Dottori 1997). Alternatively, \Op\ can be estimated by matching observations to models as suggested by Athanassoula (1992) and successfully applied by Salo et al. (1999). A valuable technique for the determination of \Op\ was introduced by Tremaine \& Weinberg (1984, hereafter referred to as the TW method), which allows the measurement of this parameter from kinematic measurements without adopting any specific dynamical model. Its application requires only measurements of the distribution of intensity and velocity of a component that reacts to the density wave. It relies on the validity of the continuity equation and assumes that the component rotates in a unique and well defined pattern. Thus the method has been mostly applied to stellar kinematic measurements (e.g., Merrifield \& Kuijken 1995; Corsini, Debattista \& Aguerri 2003; Maciejewski 2006). The TW method is, however, in principle applicable for any tracer of a component's surface density inferred from the intensity of the tracer (Zimmer, Rand, \& McGraw 2004). The main uncertainty of the TW method results is due to the presence of dust, although as shown by Gerssen \& Debattista (2007), these uncertainties do not significantly influence the derived pattern speed values. The applicability of the TW method to \Ha\ emission was shown convincingly by Hernandez et al. (2005) for M~100 and Emsellem et al. (2006) for NGC~1068 which also showed some evidence of mode coupling between different resonances. \looseness-2 

Here, we apply the TW method to \FP\ observations of the \Ha-emitting gas in the nearby late-type spiral galaxy NGC~6946, with an inclination of $38\deg$ (Carignan et al. 1990; Zimmer et al. 2004; Boomsma 2007) and at a distance of 5.5 Mpc (Kennicutt et al. 2003). NGC~6946 is a grand-design barred spiral galaxy which shows three main gravitational distortions: a large oval with radius $R\approx4.5$\arcmin, a primary bar with ellipticity 0.15 and radius $R\approx60$\arcsec, a nuclear bar with ellipticity 0.4 and radius $R\approx8$\arcsec, which is almost perpendicular to the primary bar (Elmegreen, Elmegreen, \& Montenegro 1992; Regan \& Vogel 1995; Elmegreen, Chromey, \& Santos 1998; Schinnerer et al. 2006). The disturbed spiral-arm morphology has led to the conclusion that an $m=3$ distortion is superimposed on the primary $m=2$ mode, both with the same corotation radius, $R\approx4.5$\arcmin\  (Elmegreen et al. 1992; Crosthwaite 2001). These structural properties make NGC~6946 an ideal laboratory for a detailed study of the \Op, to investigate whether the large-scale oval is capable of driving the formation and evolution of the inner components by resonance interaction. We present the data in section~\ref{sec:reduction}, followed by the analysis outline and the presentation of the results in section~\ref{sec:analysis}, and finally the discussion in section~\ref{sec:discussion}.\looseness-2

\begin{figure}
  \includegraphics[width=.5\textwidth]{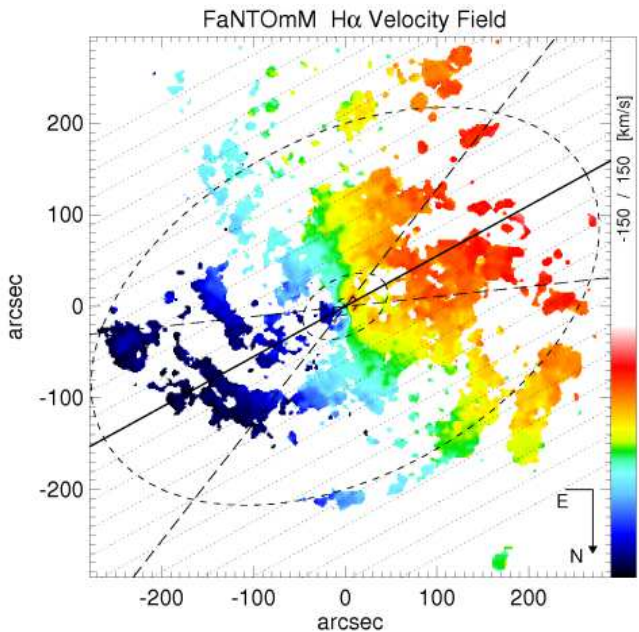}
  \caption{The observed \Ha\ velocity field for NGC~6946. The thick diagonal line shows the kinematic major-axis, and the dashed ellipses show the location of the corotation radius (outermost), the OILR (middle), and the IILR radius of the oval (see section~\ref{sec:discussion}). The dashed lines indicate the arcs inside which the rotation curve of Fig.~\ref{fig:rotcurve} has been averaged, and the parallel dotted lines show the position of blocks of 10 apertures along which velocity measurements have been extracted. The position of the dynamical center is set to $(x,y)=(0,0)$, and the systemic velocity \Vsys=46 \kms.}
  \label{fig:velfield}
\end{figure}

\begin{figure}
  \includegraphics[width=.5\textwidth]{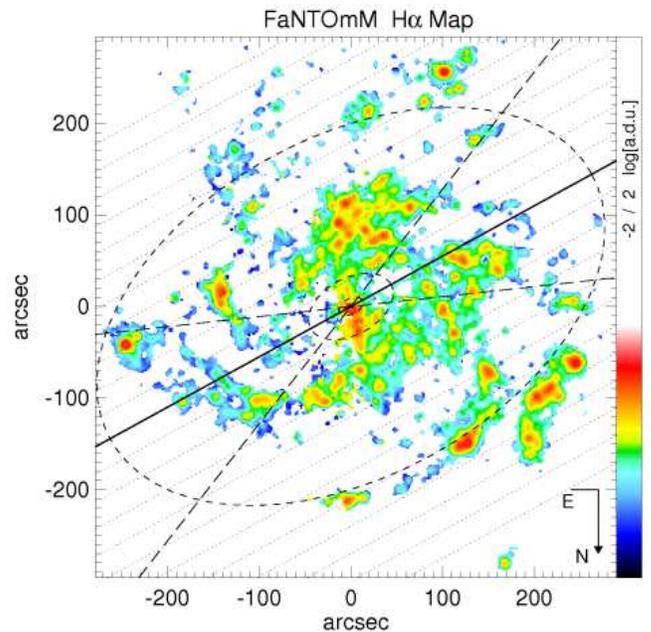}
  \caption{The observed \Ha\ intensity map used in conjunction with the velocity field presented in Fig.~\ref{fig:velfield} to calculate the \Op.}
  \label{fig:intensity}
\end{figure}

\section{Observations and Data Reduction}
\label{sec:reduction}
NGC~6946 was observed with the \fantomm\ \FP\ interferometer \citep{gachetal2002} mounted on the 1.6 meter telescope at the Observatoire du mont M\'egantic. The interference filter, centered at $\lambda_c=6569$ \AA, was used to scan the \Ha\ emission-line in 40 channels of 0.21 \AA\ during a total of 120 minutes. The \FP\ 
was tuned to a free spectral range of 391.9 \kms, over the $13.7\arcmin\times13.7\arcmin$ effective field of view with 1\farcs6/pix. Complete observation logs can be found in \citet{daigleetal2006a}. The reduction of the data cubes was performed using the \adhoc\ package \citep{amrametal1998} with consistency checks by applying the IDL-based reduction package developed by \citet{daigleetal2006b}. Further details on the reduction procedure can be found in \citet{fathietal2007}. After reduction and final cleaning, 29191 independent velocity measurements remain (see Fig.~\ref{fig:velfield}).\looseness-2

\begin{figure}
  \centering
  \includegraphics[width=.5\textwidth]{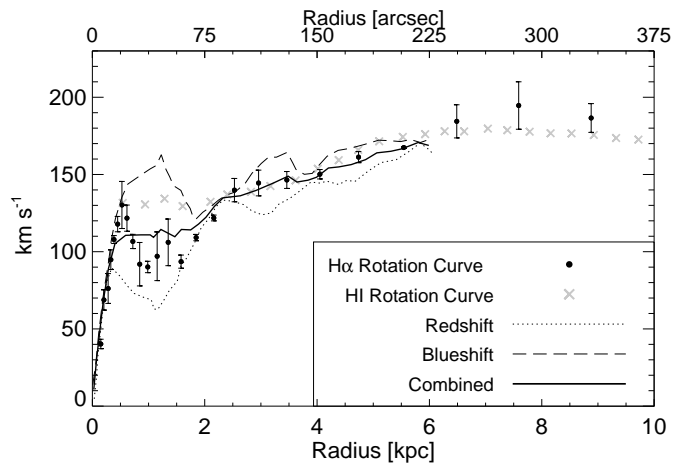}
  \caption{Our derived rotation curve (dots with error bars), the average of velocities inside the area outlined by dashed lines in Fig.~\ref{fig:velfield} (curves), and the HI rotation curve taken from \citet{boomsma_thesis} marked as crosses.}
  \label{fig:rotcurve}
\end{figure}

\section{Analysis and Results}
\label{sec:analysis}

In Cartesian coordinates, where the $x$ and $y$ axes are aligned parallel to the apparent major and minor axes of the galactic disk at inclination, $i$, the TW method yields the pattern speed 
\begin{equation}
\label{eq:TWoriginal}
\Omega_{p}  \, \sin i  = \frac{\displaystyle \int_{-\infty}^{\infty} I(x) \, [ V_\mathrm{los} (x) - V_\mathrm{sys} ] \; dx}{\displaystyle \int_{-\infty}^{\infty} I(x) \, [x - x_0] \; dx}, 
\end{equation}
where $x$ is the position along the major-axis, $I(x)$ is the observed intensity, \Vlos\ is the observed line-of-sight velocity and \Vsys, the systemic velocity. The nucleus of the galaxy is located at $x_0$, and the integrals are performed along any cut parallel to the apparent major-axis. As originally noted by \citet{TW1984}, the use of an erroneous dynamical center could introduce errors in the derivation of the \Op, although this could be avoided by using an appropriate weight function to eliminate or reduce the anomalies. \citet{MK1995} refined this technique by normalizing the measured quantities with the total intensity in the same aperture, and obtained
\begin{equation}
\label{eq:TWrefined}
\Omega_{p}=\frac{1}{\sin i}\; \frac{\langle V(x)\rangle}{\langle x\rangle},
\end{equation}
where $\langle V(x) \rangle $ is the intensity-weighted \Vlos-\Vsys\ and $ \langle x \rangle$ the intensity-weighted position of the tracer along the aperture in question. Thus the calculated \Op\ for each aperture does not change, but what changes in equation (\ref{eq:TWrefined}), is the position of each aperture in the plot of $\langle V(x)\rangle$ versus $\langle x \rangle$. As a result, the statistical importance of each aperture is evened out, and singularities in positions where $\langle x \rangle$ tends to zero, i.e., near the center of the galaxy, are avoided. \looseness-2

To derive \Op, the TW method relies on accurate derivation of the geometry of the galactic disk where the bar resides. We derive the position of the center, kinematic major-axis, and \Vsys\ by applying the tilted-ring method on the observed \Ha\ velocity field using Tukey's bi-weight mean formalism \citep{MT1977}, which is known to be non-sensitive to outliers. Applying this formalism has proved to be advantageous for cases where outliers, created by bad pixels or strong non-circular motions, can bias the derived parameters, and thus assigning smaller weights to these pixels delivers trustable disk geometry as well as kinematic parameters \citep{fathietal2005}. We derive \Vsys = $46\pm2$ \kms, and the major-axis at $241\pm2\deg$, both in good agreement with the comprehensive neutral Hydrogen analysis by \citet{carignanetal1990} and \citet{boomsma_thesis}.\looseness-2 

\begin{figure}
  \centering
  \includegraphics[width=.5\textwidth]{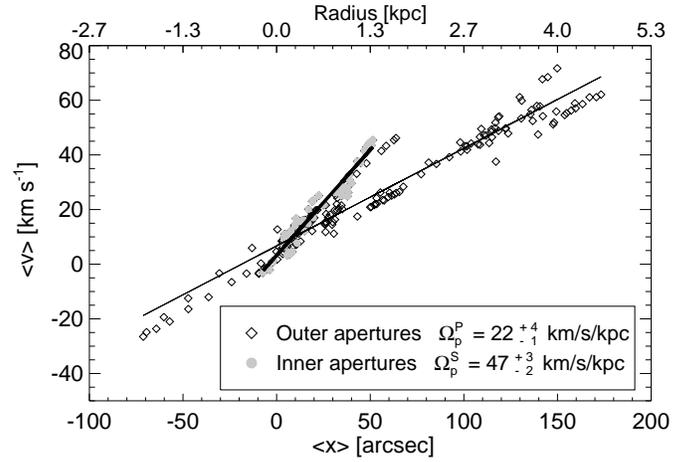}
  \caption{The TW method applied to the velocity field presented in Fig.~\ref{fig:velfield} by plotting the averaged intensity weighted velocities, $\langle V(x) \rangle $, versus the averaged intensity weighted positions, $ \langle x \rangle$, for each three-pixels wide aperture, i.e., $1/10$th of the space between the parallel dotted lines in Fig.~\ref{fig:velfield}. Open symbols show data points from apertures outside the OILR of the large oval, and filled symbols show the data points inside this radius.}
  \label{fig:TW}
\end{figure}

For the derived disk geometry and \Vsys, the corresponding rotation curve is illustrated in Fig.~\ref{fig:rotcurve}, where cuts along the redshifted and blueshifted sections of the disk as well as average rotation curve within the bounded section shown in Fig.~\ref{fig:velfield}, are overplotted. Figure~\ref{fig:rotcurve} demonstrates that the derived disk geometry results in a symmetric cut in the velocity field. Assuming a symmetric barred structure and a homogeneous dust distribution, a correct dynamical center should imply that the rotation curve extracted along the major axis is identical to that derived from the tilted-ring method. Moreover, the non-circular velocities at the blueshifted and red-shifted parts of the disk, should fall and rise with respect to the rotation curve in a symmetric fashion. This is exactly what we observe in Fig.~\ref{fig:rotcurve}. Furthermore, we note that our rotation curve is in agreement with that derived from HI studies by \citet{carignanetal1990} and \citet{boomsma_thesis}, with the advantage that our data allows us to resolve the inner regions. \looseness-2

\begin{figure}
  \centering
  \includegraphics[width=.5\textwidth]{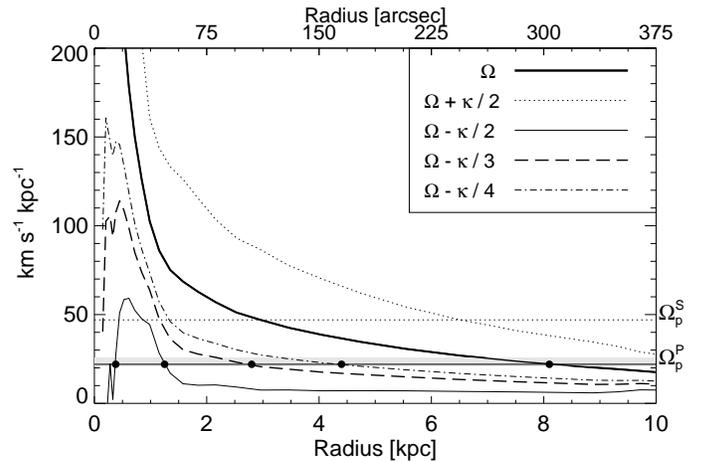}
  \caption{The angular frequency curve (thick curve) derived from the combined \Ha\ rotation curve (inner 9 kpc) and HI rotation curve (outer kpc) presented in Fig.~\ref{fig:rotcurve}. Applying the epicyclic approximation, we also plot the epicyclic frequency curves, $\Omega - \kappa/ m$ for $m=2,3,4$, and $\Omega + \kappa/ 2$. The pattern speed of the oval, with errors, is marked as the horizontal line, and the location of each resonance radius is marked with a dot. The two innermost dots mark the IILR and OILR radii discussed in section~\ref{sec:discussion}.}
  \label{fig:omega}
\end{figure}

\begin{table}
  \caption{Different pattern speed and corotation radius values for NGC~6946. 
1- \citet{Peton1982}; 
2- \citet{bonnareletal1988};  
3- \citet{ECS1998}; 
4- \citet{lpc_thesis}; 
5- \citet{zrm2004}; 
6- \citet{schinnereretal2006}; 
7- This study.} 
  \label{tab:studies}
  \begin{center}
    \leavevmode
    \begin{tabular}{lcc} \hline \hline              
Method & \Op\ (\kmskpc)& Corotation (\arcsec) \\ \hline
1- Morphology		& 60		& 78		\\
2- Morphology		& 56		& 120		\\
3- Morphology		& ---		& 155		\\
4- Morphology		& $66\pm 7$	& $300 \pm 60$	\\
5- TW on CO		& $39 \pm 13$	& ---		\\
6- Dynamical Model	& --- 		& $270 \pm 30$	\\[0.5mm]
\bf 7- TW, Primary \Opp	&\bf 22$^{+4}_{-1}$ & \bf 304$^{+13}_{-56}$ \\[0.5mm]
\bf 7- TW, Secondary \Ops&\bf 47$^{+3}_{-2}$ & ---	\\[0.5mm]
\hline 
    \end{tabular}
  \end{center}
\end{table}

To derive \Op, we set the origin of the coordinate system at the derived dynamical center (marked in Fig.~\ref{fig:velfield}) and extract the \Vlos\ along successive simulated slits parallel to the major-axis, in three pixel intervals. Thus the derived velocities within each simulated slit ($1/10$th of the space between the parallel dotted lines in Fig.~\ref{fig:velfield}) is collapsed into one single $\langle V(x)\rangle$ value, with $\langle x\rangle$  being its intensity-weighted distance along the slit. Alternatively, the spectra within each simulated slit could be collapsed into one single spectrum from which one single $\langle V(x)\rangle$ value can be derived. Both methods yield similar results. Plotting these numbers using equation (\ref{eq:TWrefined}), in Fig.~\ref{fig:TW}, it is clear that a single pattern speed cannot be assigned to all data points. Although this trend can also be seen in a similar analysis by \citet{zrm2004}, these authors do not mention or quantify it. This is probably due to the small number of points in their $\langle V(x) \rangle $ versus $ \langle x \rangle$ plot (40 data points compared with our 221). We investigate this behavior in an iterative way. \looseness-2 

Fitting all the data points, we find an average pattern speed of 21 \kmskpc. We use this initial value to analyze the angular frequency curve shown in Fig.~\ref{fig:omega}, where we find that the outer Inner Lindblad Resonance radius (OILR) of the large oval is located at around $\approx 50$\arcsec. Using this diagnosis as verified by the Kolmogorov-Smirnov test, we find that the points for apertures inside this radius (marked as gray points) fall on a unique slope clearly different from the outer apertures. We thus derive two distinct pattern speeds, the primary patter speed, \Opp=$22^{+4}_{-1}$ \kmskpc, and the secondary pattern speed \Ops=$47^{+3}_{-2}$ \kmskpc\ for the region inside the OILR. The errors include errors for the derived \Vsys\ and position angle. Using this new value for the pattern speed of the oval (compared with 21 \kmskpc) does not significantly change the position of the OILR, and thus we confirm that the inner $49$\arcsec, i.e., the region inside the OILR of the oval rotates at a rate different from that of the large oval structure. Our numerous tests, changing the width of the apertures do not induce a significant change in the derived pattern speeds, however, we note that for the inner apertures, $\langle V(x)\rangle$ has some contribution from the velocities outside the OILR of the oval. Removing all pixels outside the OILR, the secondary pattern speed could increase by 50\%.\looseness-2


\section{Discussion}
\label{sec:discussion}
In Fig.~\ref{fig:omega} we find a number of resonances, all confirming previous findings \citep{EEM1992,ECS1998}. A direct comparison between our corotation radius and previous studies is presented in table~\ref{tab:studies}, showing a good agreement between our result and those by \citet{lpc_thesis} and \citet{schinnereretal2006}. Most morphological studies yield underestimates \citep{Peton1982,bonnareletal1988,ECS1998}.  \looseness-2

The disturbed morphology of the spiral arms in NGC~6946, has for more than a decade raised the issue of the presence of $m=3$ and 4 perturbations, all following the same pattern speed as the main $m=2$ mode \citep[see][ for a detailed discussion]{schinnereretal2006}. Our data confirms the presence of two ultraharmonic resonance radii attributed to these higher modes. The $m=2$ mode, however, induces an inner Inner Lindblad Resonance radius (IILR), an OILR, and an Outer Lindblad Resonance radius. The derived OILR radius is of the same order of the primary bar, and the IILR radius falls just outside the nuclear bar, $R\approx8$\arcsec (see table~\ref{tab:resonances}).  \looseness-2

Morphological under-estimation of the corotation radius yield an over-estimated value for \Op, by almost a factor two. The study that best matches our derived \Op\ is that by \citet{zrm2004}, who apply the TW method to CO observations. We confirm that downgrading the spatial resolution of our data and deriving one unique pattern speed, we get a value comparable with that derived by \citet{zrm2004}. Our higher resolution data allow a more detailed diagnosis of the dynamics of the large-scale oval in NGC~6946, by allowing us to derive two distinct \Op\ values.  \looseness-2

\begin{table}
  \caption{The position of resonance radii with one-$\sigma$ errors.} 
  \label{tab:resonances}
  \begin{center}
    \leavevmode
    \begin{tabular}{ll} \hline \hline 
Corotation radius	       & $304^{+13}_{-56}$ arcsec $\approx$ 8.30 kpc\\[0.9mm]
$\Omega-{\kappa \over 4}$      & $165^{+11}_{-38}$ arcsec $\approx$ 4.40 kpc\\[0.9mm]
$\Omega-{\kappa \over 3}$      & $105^{+6}_{-34}$ arcsec $\approx$ 2.80 kpc\\[0.9mm]
OILR			       & $49^{+2}_{-4}$ arcsec $\approx$ 1.25 kpc\\[0.9mm]
IILR			       & $14^{+1}_{-1}$ arcsec $\approx$ 0.38 kpc\\[0.9mm]
\hline 
    \end{tabular}
  \end{center}
\end{table}

This observational result confirms a ``bars within bars within bars'' scenario, where an evolved large primary $m=2$ perturbation forms an inner bar whose pattern speed could be larger by a factor of up to three \citep[e.g., ][]{MS2000,ES2004}. The bar may be found at any position angle with respect to the large scale oval, and such systems are known to form nuclear bars \citep[e.g.,][]{ES1999}, suggesting that multiple bars probably generally have different pattern speeds. In Fig.~\ref{fig:omega}, we see that \Ops\ is consistent with the nuclear bar being located at the ILR of the primary bar in NGC~6946. Furthermore, as found by \citet{ECS1998}, the nuclear bar is almost perpendicular to the larger bar, which they interpret as the presence of the well known $x_1$ and $x_2$ orbits of the same structure. Placing the corotation of the larger bar at the OILR of the oval agrees with the nuclear bar being inside the ILR of the primary bar, i.e., the IILR of the oval structure. Such a scenario is also known to drive efficient mass flow from the outer parts toward the nuclear regions in galaxies (Shlosman, Frank, \& Begelman 1989) where a starburst can occur \citep[e.g., ][]{zuritaetal2004,fathietal2005}. \looseness-2

The tight agreement between the sizes of the different components with the predicted location of the resonances confirm that the inner components have formed by means of resonance interaction of the main $m=2$ gravitational perturbation, i.e., the large oval and two prominent spiral arms. As discussed for example by Maciejewski (2004), linear theory is applicable only if the gravitational perturbation is not strong. In NGC~6946, this is confirmed since the main perturbation is an oval distorsion and not a strong bar. The inflow is confirmed by our residual velocity field after subtracting the rotational model from the observed velocity field. A detailed discussion is beyond the scope of this letter, however, the velocity field presented in Fig.~\ref{fig:velfield} shows the expected signature of the OILR as a ``kink'' in the contour tracing the systemic velocity. \looseness-2

\begin{acknowledgements}
We thank the referee for insightful comments which helped improve this manuscript. KF acknowledges support from the Wenner-Gren foundations and the Royal Swedish Academy of Sciences' Hierta-Retzius foundation. KF and JEB acknowledge support through the IAC project P3/86 as well as the Spanish Ministry of Educational Science grant, AYA2004-08251-CO2-01. ST wishes to thank the IAC for  hospitality during a 4 month visit. \looseness-2
 \end{acknowledgements}


\end{document}